\documentstyle[12pt]{report}
\begin{document}

\thispagestyle{empty}

\vspace*{1cm}

{\Huge
\bf
\noindent
%\begin{center}
Quantum Hall Systems\\[10pt] 
%\end{center}
}
\vspace{1cm}

\noindent
{\large
\bf
Braid groups, composite fermions,
and fractional charge
}

\vspace{1cm}

{\large
\it

\noindent
\hspace*{3cm} Lucjan Jacak\\
\hspace*{3cm} Piotr Sitko\\
\hspace*{3cm} Konrad Wieczorek\\
\hspace*{3cm} Arkadiusz W\'ojs

\vspace{3cm}

}

{\rm
\begin{center}
%\noindent
Institute of Physics\\
Wroc\l{}aw University of Technology\\
Wroc\l{}aw, Poland

\end{center}

}

\newpage
\thispagestyle{empty}

\vspace*{11cm}

\paragraph*{Acknowledgement}
{\it The Authors would like to acknowledge Professor John J. Quinn 
as coauthor of original ideas of chapters 8 and 9.}

\newpage

\setcounter{page}{3}
\thispagestyle{empty}
\vspace{20pt}

\noindent
{\Huge\bf Table of contents}\\[40pt]

%\vspace{1cm}

\noindent {\bf 1. Introduction} \dotfill %6
\\

\noindent {\bf 2. Topological
methods for description\\ 
\hspace*{0.5cm}  of quantum many-body systems}\dotfill %21
\\[12pt]
\hspace*{0.5cm} 2.1. Configuration spaces of quantum
many-body systems\\
\hspace*{1.3cm} of various dimensions\dotfill
%21
\\[4pt]
\hspace*{0.5cm} 2.2. Quantization of many-body systems\dotfill
%25
\\[4pt]
\hspace*{0.5cm} 2.3. The first homotopy group for   
the many-particle \\ 
\hspace*{1.3cm} configuration space -- braid groups\dotfill %27
\\[4pt]
\hspace*{0,5cm} 2.4. Braid groups for specific manifolds\dotfill
%30
\\

\noindent {\bf 3. Quantization
of many-particle systems\\ 
\hspace*{0.5cm}   and quantum 
statistics in lower dimensions}\dotfill %42
\\[12pt]
\hspace*{0.5cm} 3.1. Topological limitations of 
quantum-mechanical description\\
\hspace*{1.3cm} of many-particle systems  \dotfill %42
\\[4pt]
\hspace*{0.5cm} 3.2. Quantum statistics and 
irreducible unitary representations
\\
\hspace*{1.3cm} of braid groups for selected manifolds \dotfill %43
\\[4pt]
\hspace*{0.5cm} 3.3. Non-Abelian statistics \dotfill %53
\\[4pt]

\noindent {\bf 4. Topological approach to 
composite particles\\
\hspace*{0.5cm} in two dimensions}\dotfill %56
\\[12pt]
\hspace*{0.5cm} 4.1. Mathematical model of composite particles\dotfill %56
\\[4pt] 
\hspace*{0.5cm} 4.2. Configuration space for the system
of composite particles\dotfill %63
\\[4pt] 

\noindent {\bf 5. Many-body methods for  Chern--Simons 
systems}\dotfill %71
\\[12pt]
\hspace*{0,5cm} 5.1. Random phase approximation for an anyon gas\dotfill
%71
\\[4pt]
\hspace*{0,5cm} 5.2. Correlation energy of an anyon gas\dotfill %77
\\[4pt]
\hspace*{0,5cm} 5.3. Hartree--Fock approximation for  
Chern--Simons systems\dotfill %80
\\[4pt]
\hspace*{0,5cm} 5.4. Diagram analysis for a gas of anyons\dotfill %84
\\

\thispagestyle{empty}

\noindent {\bf 6. Anyon superconductivity}\dotfill %91
\\[12pt]
\hspace*{0,5cm} 6.1. Meissner effect  in an  anyon gas at $\rm T=0$\dotfill
%91
\\[4pt]
\hspace*{0,5cm} 6.2. Gas of anyons at finite temperatures\dotfill %95
\\[4pt]
\hspace*{0,5cm} 6.3. Higgs mechanism in an anyon superconductor\dotfill
%98
\\[4pt]
\hspace*{0,5cm} 6.4. Ground state of an anyon superconductor in the
Hartree--Fock\\ 
\hspace*{1.3cm} approximation\dotfill %102
\\

\noindent {\bf 7. The fractional quantum Hall effect\\
\hspace*{0.5cm}  in composite fermion systems}\dotfill %105
\\[12pt]
\hspace*{0,5cm} 7.1. Hall conductivity in a system of composite fermions\dotfill
%105
\\[4pt]
\hspace*{0,5cm} 7.2. Ground state energy of composite fermion systems\dotfill
%111
\\[4pt]
\hspace*{0,5cm} 7.3. Metal of composite fermions\dotfill %114
\\[4pt]
\hspace*{0,5cm} 7.4. BCS paired Hall state\dotfill %117
\\

\noindent {\bf 8. Quantum Hall systems on a sphere}\dotfill 
%121
\\[12pt]
\hspace*{0,5cm} 8.1. Spherical system\dotfill
%121
\\[4pt]
\hspace*{0,5cm} 8.2. Composite fermion transformation \dotfill
%127
\\[4pt]
\hspace*{0,5cm} 8.3. Hierarchy\dotfill
%131
\\

\thispagestyle{empty}

\noindent {\bf 9. Pseudopotential approach to the fractional quantum\\
\hspace*{0.5cm} Hall states}\dotfill %137
\\[12pt]
\hspace*{0.5cm} 9.1. Problems with justification of the composite \\
\hspace*{1.3cm} fermion picture \dotfill %137
\\[4pt]
\hspace*{0.5cm} 9.2. Numerical studies on the Haldane sphere\dotfill %139
\\[4pt]
\hspace*{0.5cm} 9.3. Pseudopotential approach \dotfill %142
\\[4pt]
\hspace*{0.5cm} 9.4. Energy spectra of short range pseudopotentials 
\dotfill %144
\\[4pt]
\hspace*{0.5cm} 9.5. Definition of short range pseudopotential 
\dotfill %147
\\[4pt]
\hspace*{0.5cm} 9.6. Application to various pseudopotentials
\dotfill %149
\\[4pt]
\hspace*{0.5cm} 9.7. Multi-component systems \dotfill %154
\\[4pt]

\noindent {\bf Appendix A. Homotopy groups}\dotfill %160
\\

\noindent {\bf Appendix B. Correlation function for anyon gas\\
\hspace*{2.7cm} in the self-consistent Hartree approximation}\dotfill %164
\\

\noindent {\bf Index}\dotfill %169
\\

\noindent {\bf References}\dotfill %172
\\

\thispagestyle{empty}

\noindent
A sample  of this book is available at:\\
http://www.opu.co.uk/isbn/0-19-852870-1

\end{document}